\documentclass[aip,twocolumn,english,superscriptaddress,preprintnumbers,amsmath,amssymb,showpacs]{revtex4}
\usepackage[T1]{fontenc}
\usepackage[latin9]{inputenc}
\usepackage{amsmath}
\usepackage{graphics}
\usepackage{graphicx}
\usepackage{epsfig}
\usepackage{amssymb}
\usepackage{dcolumn}
\usepackage{bm}
\usepackage{float}
\usepackage{leftidx}
\usepackage{booktabs}
\makeatletter
\@ifundefined{textcolor}{}
{%
 \definecolor{BLACK}{gray}{0}
 \definecolor{WHITE}{gray}{1}
 \definecolor{RED}{rgb}{1,0,0}
 \definecolor{GREEN}{rgb}{0,1,0}
 \definecolor{BLUE}{rgb}{0,0,1}
 \definecolor{CYAN}{cmyk}{1,0,0,0}
 \definecolor{MAGENTA}{cmyk}{0,1,0,0}
 \definecolor{YELLOW}{cmyk}{0,0,1,0}
 }

\makeatother

\usepackage{babel}

\begin{document}

\title{Simulations of stable compact proton beam acceleration from a two-ion-species ultrathin foil}

\author{T.~P. Yu}
\thanks{Electronic mail: Tongpu.Yu@tp1.uni-duesseldorf.de.}
 \affiliation{Institut f\"ur Theoretische Physik I, Heinrich-Heine-Universit\"at D\"usseldorf, 40225 D\"usseldorf, Germany}
 \affiliation{Department of Physics, National University of Defense Technology, Changsha 410073, China}

\author{A. Pukhov}
 \thanks{Electronic mail: pukhov@tp1.uni-duesseldorf.de.}
 \affiliation{Institut f\"ur Theoretische Physik I, Heinrich-Heine-Universit\"at D\"usseldorf, 40225 D\"usseldorf, Germany}

\author{G. Shvets}
\thanks{Electronic mail: gena@physics.utexas.edu.}
 \affiliation{The University of Texas at Austin, Department of Physics, Austin, TX 78712}

\author{M. Chen}
 \affiliation{Institut f\"ur Theoretische Physik I, Heinrich-Heine-Universit\"at D\"usseldorf, 40225 D\"usseldorf, Germany}
  \affiliation{Accelerator $\&$ Fusion Research Division, Lawrence Berkeley National Laboratory, Berkeley, California 94720, USA}

\author{T.~H. Ratliff}
 \affiliation{The University of Texas at Austin, Department of Physics, Austin, TX 78712}

\author{S.~A. Yi}
 \affiliation{The University of Texas at Austin, Department of Physics, Austin, TX 78712}

\author{V. Khudik}
 \affiliation{The University of Texas at Austin, Department of Physics, Austin, TX 78712}

\date{\today}


\begin{abstract}
We report stable laser-driven proton beam acceleration from ultrathin foils consisting of two ion species:
heavier carbon ions and lighter protons. Multi-dimensional particle-in-cell (PIC) simulations show that the radiation
pressure leads to very fast and complete spatial separation of the species. The laser pulse does not penetrate the carbon
ion layer, avoiding the proton Rayleigh-Taylor-like (RT) instability. Ultimately, the carbon ions are heated and spread
extensively in space. In contrast, protons always ride on the front of the carbon ion
cloud, forming a compact high quality bunch. We introduce a simple three-interface model to interpret the instability suppression
in the proton layer. The model is backed by simulations of various compound foils such as carbon-deuterium (C-D) and carbon-tritium (C-T) foils.
The effects of the carbon ions' charge state on proton acceleration are also investigated.
It is shown that with the decrease of the carbon ion charge state, both the RT-like instability and the Coulomb explosion degrade the energy spectrum of the protons.
Finally, full 3D simulations are performed to demonstrate the robustness
of the stable two-ion-species regime.
\end{abstract}



\maketitle

\section{\label{sec:level1}INTRODUCTION\protect}

In recent years, laser-plasma based ion accelerators have drawn increased attention due to their
unique features such as compact size, low beam
emittance, high accelerating gradient, and high peak energy. Several acceleration mechanisms
have been proposed to produce such ion beams, for example, target normal sheath
acceleration (TNSA)~\cite{s1a,s1b,s1c,s1d,s1e}, shock wave acceleration~\cite{s2a,s2b,s2c},
Coulomb explosion~\cite{s3a}, and the break-out afterburner (BOA)~\cite{s4}.
In such schemes with laser intensity less than $10^{20}W/cm^2$, ions within a few wavelengths of the
foil surface are accelerated by the charge separation
field $E_{s}$ at the target rear, and shock wave acceleration or hole boring~\cite{s44} at the target
front. If the laser pulse is intense enough to evacuate
all the electrons from the focal spot via $J\times B$ force, the ion acceleration is
usually accompanied by Coulomb explosion or even the thermal expansion~\cite{s45}.
In these cases, a linearly polarized laser is usually preferable due to its ability
to heat the electrons through the oscillating part of its ponderomotive force.

However, both the beam quality and the energy conversion efficiency of these mechanisms are far away
from the requirements of real applications. The inefficiency and lack of tunability
probably come from the fact that the acceleration is only due to space charge effects. Recently,
with the rapid development of plasma mirror technology~\cite{s5}, both the laser intensity (in excess of $10^{22}W/cm^{2}$) and
the contrast ($\sim10^{-12}$) have been increased significantly. One of the most straightforward acceleration mechanisms,
radiation pressure acceleration~\cite{s6a,s6b,s6c,s6d,s6e,s6g} (RPA, also called laser piston~\cite{s6a}
or light sail~\cite{s6e}), is being re-visited. Circularly polarized laser pulses are
used to reduce the electron heating by directly pushing the electrons inward into the target~\cite{s7}.
In the fist stage of the RPA regime, the plasma foil works like a rectifier to convert the transverse
laser field to a strong longitudinal accelerating field. The radiation pressure, $P=2I/c$,
exerted by the intense laser pulse is up to several tens of Gbar and accelerates forward the entire irradiated region of the foil, where $I=2a_{0}^{2}\times 1.37\times10^{18}/(\lambda[\mu m])^{2}(W/cm^{2})$
is the laser intensity, $c$ the light speed, $a_{0}$ the laser dimensionless electric field
and $\lambda$ the laser wavelength. In principle, unlimited ion energy gain is obtainable with a sufficiently long pulse.

Much work has been done to investigate the RPA mechanism. A semi-analytical RPA model~\cite{s6a,s6b} was proposed which is
verified by a series of particle-in-cell (PIC) simulations. Typically, a "spiral structure"~\cite{s6d,s7} in phase space is observed,
resulting in a mono-energetic spectrum with a high energy conversion efficiency.
The first RPA experiment~\cite{s8}, performed recently, also observed phase-stable~\cite{s6d} ion acceleration from laser radiation pressure. However, several issues
must be solved before the scheme is ready for practical applications. One of the most important
issues is the fast growth of an undesirable transverse instability, similar to the
Rayleigh-Taylor (RT) instability in inertial confinement fusion~\cite{s6a,s9a}.
Some schemes have been suggested to suppress the RT-like instability, for example,
using a properly tailored laser pulse. However, in most schemes the RT instability inevitably tears the foil into many bubbles and clumps\cite{s10}.
Though the ions' energy distribution is initially monoenergetic,
it degrades to a quasi-exponential distribution after several laser cycles. In Ref.~\cite{s11}, we proposed a new regime
of stable proton beam acceleration by using a two-ion-species foil.
Two-dimensional (2D) PIC simulations have shown that the two ion species from an initially homogeneous hydrocarbon target can separate into distinct layers at the beginning of the interaction. The carbon and proton layers continue to share an interface, and are accelerated together.
The high quality proton beam persists after the laser-foil interaction ends.

In this paper, we provide insight into the influences of the foil composition
and the laser penetration into the foil on the ion acceleration. We find that the scheme using two-ion-species
shaped foils effectively avoids the foil deformation and significantly suppresses
the RT-like instability in the proton layer, if the laser intensity and the foil composition are appropriately chosen.
In the following, we firstly review the 1D RPA model and then discuss
how to extend it to multi-dimensional cases and how to further enhance the stable proton beam acceleration.

\section{RPA model and 1D simulation results}

We begin by reviewing the semi-analytical RPA model~\cite{s6a,s12}. When an ultra-intense laser
pulse irradiates an ultra-thin over-dense foil, the foil can be regarded as a plasma mirror.
Most of the laser energy is transferred to the foil with only a small part carried away by the reflected laser pulse.
Assuming the frequency of the incident and reflected laser pulse to be $\omega_{0}$ and $\omega_{1}$,
respectively, we have $\omega_{1}/\omega_{0}=(1-\beta)/(1+\beta)\simeq 1/4\gamma^{2}$ as a result of the
red-shift, where $\beta$ is the foil velocity normalized by the light speed
$c$ and $\gamma$ is the relativistic factor. As the foil is accelerated to near light speed, the laser energy is efficiently converted to kinetic energy of the foil, with energy conversion efficiency
$\eta=2\beta/(1+\beta)\simeq1-1/4\gamma^{2}\approx1$.  For an ultrathin foil,
we may assume that the accelerated region of the foil remains intact. The foil
motion can thus be described as follows:

\begin{eqnarray}
\rho{d(\gamma \beta)\over dt}={{E_L}^2\over 2\pi c}{{1-\beta}\over {1+\beta}},\label{eq:No.1}
\end{eqnarray}

\noindent where $\rho=\sum_{\substack{i}} m_in_il$ is the target areal mass density,
$m_i$, $n_i$, and $l$ are the ion mass, ion density, and foil thickness, respectively.
$E_L$ represents the laser electric field. The foil dynamics in the two-ion-species
case is defined by the areal mass density $\rho=(m_{1}n_{1}+m_{2}n_{2})l$, not the structure within the foil,
which is quite different from the collisionless shock wave acceleration in Ref.~\cite{s2b,s2c}.
By using hyperbolic functions and defining $\beta=tanh \theta$, we can get a very compact
form of the parameters $\gamma=cosh \theta$ and $\gamma\beta=p=sinh \theta$.
Then, the Eq.~\eqref{eq:No.1} can be rewritten as

\begin{eqnarray}
(cosh\theta+sinh\theta)^{2}d(sinh\theta)=Adt, A=\frac{{E_{L}}^2}{2\pi c\rho}\label{eq:No.2}
\end{eqnarray}

\noindent Noting $\sqrt{1+u}du=\frac{2}{3}d(1+u)^{\frac{3}{2}}$ and considering the initial condition
$\beta_{t=0}\sim0$, we obtain

\begin{eqnarray}
p+\frac{2}{3}p^{3}+\frac{2}{3}\gamma^{3}=At+\frac{2}{3}.\label{eq:No.3}
\end{eqnarray}

\noindent For $p\ll1$ or $\beta\ll\frac{\sqrt{2}}{2}$, $p\approx At$.
As the target approaches the speed of light ($\beta \rightarrow 1$), $p\approx (\frac{3}{4}At)^{\frac{1}{3}}$. This indicates that the ion energy initially increases at a rate of $(It/\rho)^{2}$, but slows to $(It/\rho)^{\frac{1}{3}}$. Further, we can obtain the exact solution of the Eq.~\eqref{eq:No.1} as below

\begin{eqnarray}
{\beta}={1-[\varsigma(t)+\varkappa(t)]^{\frac{1}{3}}-[\varsigma(t)-\varkappa(t)]^{\frac{1}{3}}},\label{eq:No.5}\\
{\gamma-1}={(\xi(t)-1)^2\over 2\xi(t)},\label{eq:No.6}
\end{eqnarray}

\noindent where $\varsigma(t)=1/[1+h(t)^{2}]$, $\varkappa(t)=h(t)/[1+h(t)^{2}]^{\frac{3}{2}}$,
$\xi(t)=\sqrt{(1+\beta)/(1-\beta)}$, and $h(t)=3At+2$. Eq.~\eqref{eq:No.5} is based on the assumption that the foil remains intact, i.e. both ion species are accelerated to the same velocity.

We carry out a series of 1D simulations to investigate the detailed acceleration process
by using the PIC code VLPL~\cite{s13}. We let the heavier
species be carbon ions and the lighter species be protons. In the first case,
the longitudinal length of the simulation box is $x=60 \lambda$ with $6\times 10^4$
cells so that the expected density spike can be resolved. For simplicity, we take the laser wavelength as $\lambda=1.0 \mu m$.
Each cell contains about $100$ numerical macro-particles. The target is $0.1\lambda$ long, located at $x=10\lambda$
and composed of fully ionized carbon ions and protons with the same number density $71.42n_c$ (the critical density $n_{c}\sim 1.1\times 10^{21}/cm^{3}$ for $\lambda=1.0 \mu m$),
which corresponds to an electron density $n_e=500n_c$. A circularly polarized laser pulse
is incident from the left boundary at $t=0$. The laser intensity
follows a trapezoidal profile (linear growth - plateau - linear decrease) in time.
The dimensionless laser intensity is $a_0=100$ and the duration is
$\tau_{L}=16 T_0$ ($\sim 52.8 fs, 1T_0-14T_0-1T_0$). Absorbing boundary conditions are applied to
both the fields and particles.

\begin{figure}[!htb]
\suppressfloats\includegraphics[width=8.6cm]{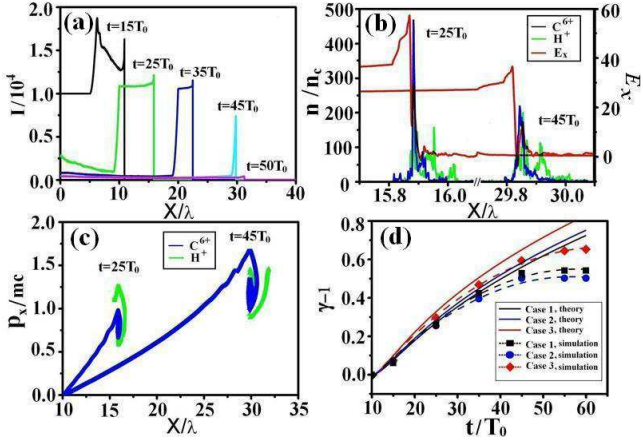}\caption{\label{f1}(Color online)
(a) Laser intensity evolution. Here, $I=(E_{y}^2+B_{z}^2)/2$ is normalized to
$I_{0}=2.74\times 10^{18} W/cm^{2}$. The laser-foil interaction ends at $t=45T_{0}$.
(b) Carbon ion and proton density distributions at $t=25T_{0}$ and $45T_{0}$.
 The red color shows the accelerating field $E_{x}$ which is normalized to $E_{0}=m_{e}c\omega/e=3.2\times10^{12}V/m$.
(c) Carbon ion and proton phase space distributions at $t=25T_{0}$ and $45T_{0}$.
(d) Proton energy evolution from 1D PIC simulations and 1D RPA model.}
\end{figure}

Fig.~\ref{f1}(a) shows the laser intensity evolution. The wave front of the laser pulse
arrives at the foil surface at $t=10T_{0}$. We can see that a part of the incident laser pulse
is initially reflected by the target at $t=15T_{0}$ because the foil is opaque to the laser.
The laser energy is transferred to the foil until the laser-foil
interaction ends at about $t=45T_{0}$. Fig.~\ref{f1}(b) presents the distribution of
the ion density $n_{C}$, $n_{H}$, and the accelerating field $E_{x}$. At the first stage of the RPA, the electrons
are pushed out by the $J\times B$ force and a strong charge separation field forms at the foil rear.
Because of the higher charge-to-mass ratio $Z_{H}/m_{H}$, protons quickly move to the front of the foil at the beginning of the interaction.
The separation time can be estimated by
$t_{sep}=\sqrt{2lm_{H}/eE_L}= 2.5fs$. The ions then experience different accelerating fields,
as shown by the red spike in the field distribution. The accelerating field inside the carbon layer
is much stronger than in the proton layer so that the carbon ions can be accelerated with
the protons. The acceleration process repeats until $t=45T_{0}$, similar to the "snow-plough" in the
electron acceleration~\cite{s14}.

Our simulation results also agree well with the Refs.~\cite{s6d,s7},
where a typical "spiral structure" was observed in a pure hydrogen foil. The fact that both the heads
of carbon ions and protons interlace with each other in phase space, as shown in Fig.~\ref{f1}(c),
demonstrates the above assumption on the ion acceleration process.

The averaged proton energy evolution is shown in Fig.~\ref{f1}(d). At $t=45T_{0}$, the proton
energy is as high as $500MeV$, which is slightly higher than the carbon ion energy $450MeV/u$.
Such high energies with a pronounced mono-energetic peak are yet unreachable in other acceleration mechanisms.
For comparison, we also show the theoretical predictions of Eq.\eqref{eq:No.6} in the figure,
which agree well with the simulation results.

\begin{figure}[!htb]
\suppressfloats\includegraphics[width=7.5cm]{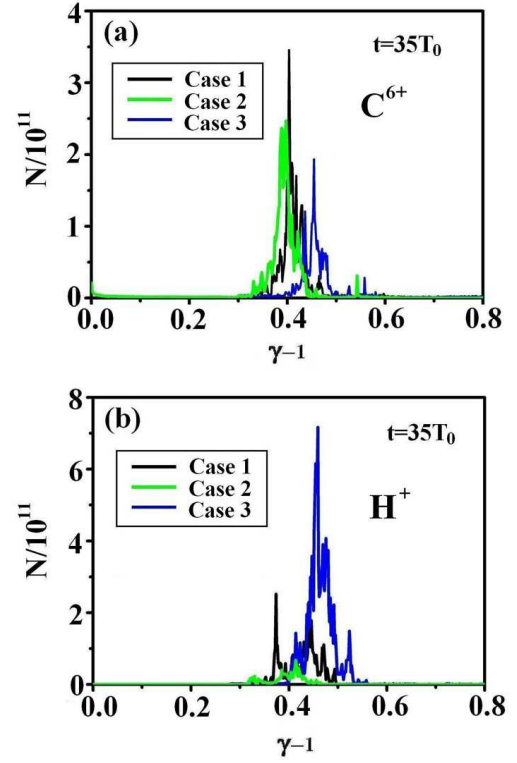}\caption{\label{f2}(Color online)
Energy spectra of (a) carbon ions and (b) protons at $t=35T_{0}$. The ion density ratio
$n_{C}:n_{H}$ in all three cases is 1:1 (case 1), 4:1 (case 2),
and 1:4 (case 3), respectively. A clear mono-energetic peak for both carbon ions
and protons is observed in all three cases.}
\end{figure}

We also performed simulations to investigate the influence of the foil composition on the final
proton energy. The laser and foil parameters are the same as in the case above, except
for the ion density ratio. We keep the electron density the same but vary the ion density ratio
$n_{C}:n_{H}$ from $1:1$ (case 1) to $4:1$ (case 2) and $1:4$ (case 3).
The areal mass density $\rho$ in the three cases is $\rho_{2}:\rho_{1}:\rho_{3}=1.06:1.00:0.86$.
The ion energy spectra at $t=35T_{0}$ are shown in Fig.~\ref{f2}. Apparently, both carbon
ions and protons show a clear mono-energetic peak in spite of the different areal mass density $\rho$.
Overall, the carbon ion energy per nucleon is a little lower than the proton's. For the case 1 and 2, they have similar ion energy spectra
due to the close areal mass density. For the case 3, the protons can be accelerated
more efficiently because of the lower areal mass density $\rho_{3}$. The averaged ion energy
evolution is also shown in Fig.~\ref{f1}(d). As expected, a similar curve for the case 1 and 2
is observed though the proton density significantly decreases from $71.42n_{c}$ to $20n_{c}$.
This demonstrates that the overall foil acceleration depends only weakly on the foil composition.
While reducing the areal mass density, the protons tend to be more efficiently accelerated.
We also find that the average proton energies for the cases are very close
to the theoretical predictions as marked by the solid curve in the figure, though a deviation is observed at the post-interaction stage.

We should mention that radiation reaction (RR) effects are generally recognized to become important when the laser intensity is above $10^{22}W/cm^{2}$~\cite{s16a,s16b}.
However, recent researches indicate that, for a circularly polarized laser pulse,
the RR effects become relevant only when the foil is thin enough for the laser pulse to break through it~\cite{s16c,s16d}. In this case,
the final proton energy can be even increased slightly. However, such effects are not the focus of this paper and are neglected.

\section{Multi-dimensional effects and shaped foil target (SFT)}
The 1D numerical simulations above reveal that the RPA dominated regime has the potential
to generate ion beams with favorable qualities such as high intensity and short duration. Mono-energetic ion beams with peak energy $200-500MeV/u$ are in demand for medical therapy of deep-seated abdominal tumors, treatment of brain cancers~\cite{s17}, and proton-driven fast ignition of fusion targets.

However, when the 1D model above is extended to multi-dimensional cases, several issues arise.
First, the foil is deformed by the incident laser pulse due to the transverse laser profile.
This leads to strong electron heating which degrades the final ion energy spectrum.
Second, the ultrathin foil is very susceptible to transverse instabilities, such as a RT-like instability~\cite{s6b, s11}.  The RT-like instability is seeded once the laser-foil
interaction begins, and develops at the unstable interface within a few laser cycles.
Gradually, the foil surface becomes corrugated and pierced by the laser radiation and the
entire target is torn into many clumps and bubbles~\cite{s10}. The final energy spectrum
then shows a quasi-exponential decay with a sharp cut-off energy. Third, the ion
acceleration in the RPA regime is usually accompanied by other acceleration mechanisms,
such as TNSA, Coulomb explosion, and even thermal expansion, making
the acceleration process complex and nonlinear.

\begin{figure}[!htb]
\suppressfloats\includegraphics[width=7.6cm]{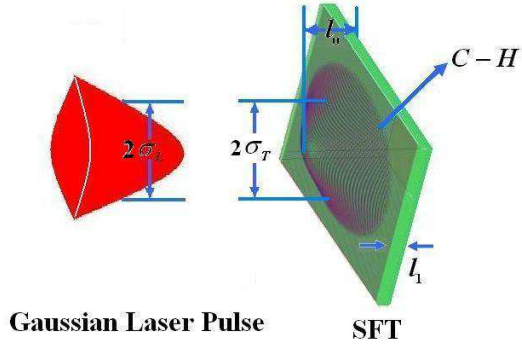}\caption{\label{f3}(Color online)
Schematic of the two-ion-spices shaped foil target (SFT). A Gaussian laser pulse is incident
from the left boundary. The profile of the foil is defined by three parameters: $l_{0}$,
$l_{1}$, and $\sigma_{T}$. The foil is composed of mixed carbon ions and protons with
various particle density ratios $n_{C}:n_{H}$.}
\end{figure}

In order to avoid foil deformation, we employ a shaped foil target (SFT)~\cite{s18a,s18b,s18c}
to compensate for the transverse laser profile, as suggested by the Eq.\eqref{eq:No.1},
i.e., $E_{L}^{2}\varpropto \rho\varpropto l$. Considering the usual
Gaussian laser pulse, for example, the transverse foil thickness $l_{y}$ follows
$l_{y}=max[l_{0}exp(-y^{2}/\sigma_{T}^{2}), l_{1}]$, where $l_{0}$ is the maximal
foil thickness, $l_{1}$ the cut-off thickness, and $\sigma_{T}$ the laser spot size,
as shown in Fig.~\ref{f3}. Such a foil geometry can be fabricated for experiments by
employing polyethylene foils~\cite{s8}.
Alternative methods include molecular beam epitaxy technique (MBE)~\cite{s24} or deposition techniques
for thin films, such as physical vapor deposition (PVD)~\cite{s24a} or chemical vapor deposition (CVD)~\cite{s24b}.
By directly matching the transverse laser profile and target areal mass density, each foil element $ l_{x}dl_{y}$ behaves like in the 1D case
and the irradiated spot can be uniformly pushed forward. A single ion species SFT irradiated by a
circularly polarized laser pulse has been numerically investigated in Refs.~\cite{s18a,s18b,s18c}, where an
improved acceleration structure was shown. However, transverse instability of the foil still degrades
the initially mono-energetic ion beam. In the following, we show 2D simulations of two-ion-species
foils, i.e., hydrocarbon SFTs, which present different dynamics from the hydrogen SFT in Ref.~\cite{s18a}.

\section{2D simulation results}

Two basic cases are first investigated with 2D simulations. In the case A, the simulation box is
$X\times Y=50\lambda\times50\lambda$, sampled by $10000\times5000$ cells.
Each cell contains $100$ numerical macro-particles in the plasma region. The foil is initially located at
$x=10\lambda$ with parameters $l_{0}=0.1\lambda$, $l_{1}=0.05\lambda$, and $\sigma_{T}=7\lambda$.
Both species have the same particle density $n_{C}=n_{H}=45.71n_{c}$, which corresponds to the electron
density $n_{e}=320n_{c}$ ($\sim3.5\times10^{23}/cm^{3}$). A circularly polarized Gaussian laser pulse with the focal size
$\sigma_{L}=8\lambda$ is incident from the left boundary. The laser duration is $\tau_{L}=10T_{0}$
$(\sim 33 fs, 1T_{0}-8T_{0}-1T_{0})$. All the other parameters are the same as in the 1D cases\ above.

\begin{figure}[!htb]
\suppressfloats\includegraphics[width=8.6cm]{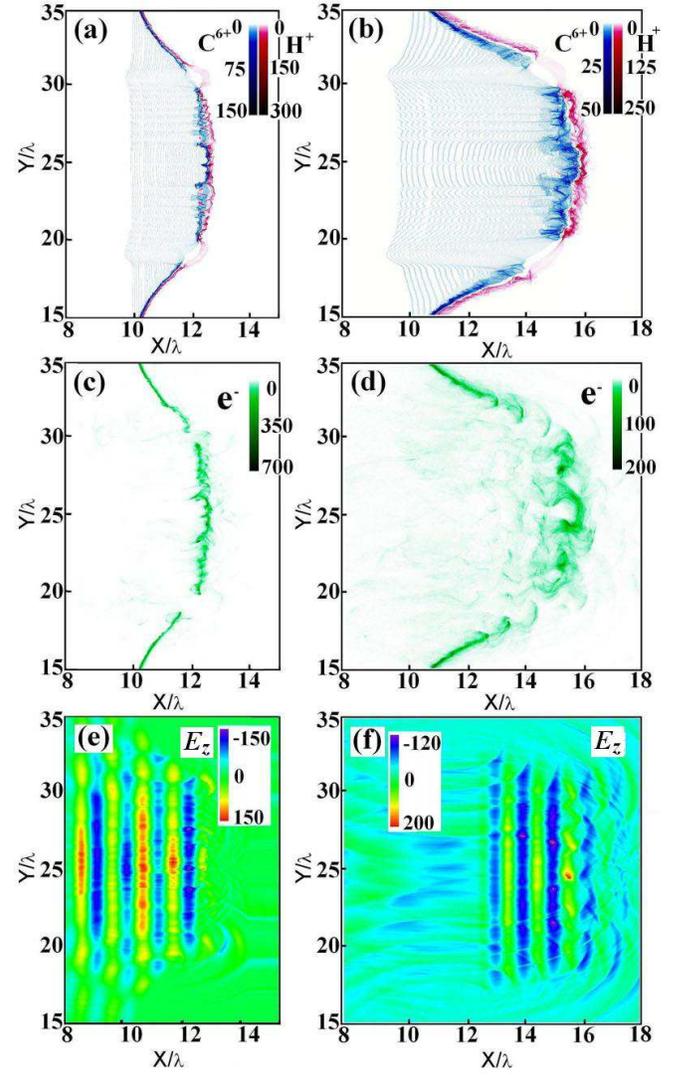}\caption{\label{f4}(Color online)
Density distributions of protons and carbon ions in the case A at (a) $t=17.5T_{0}$
and (b) $22.5T_{0}$. RT instability is observed initially for both carbon ions and protons.
The corresponding electron density distribution and the laser electric field $E_{z}$ are
shown in frames (c-d) and (e-f), respectively. Here, the density is normalized to $n_{c}$
and the field to $E_{0}$.}
\end{figure}

\begin{figure}[!htb]
\suppressfloats\includegraphics[width=8.6cm]{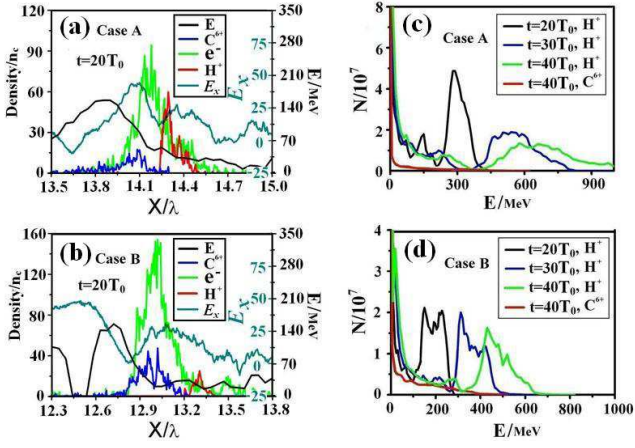}\caption{\label{f5}(Color online)
 Particle density distributions and the electron energy distribution as well as the accelerating
  fields $E_{x}$ on the laser axis at $t=20T_{0}$ in the (a) case A and (b) case B.
 Energy spectrum evolutions of the protons and carbon ions at $t=20, 30, 40T_{0}$
 in the (c) case A and (d) case B. For the case B, the compact proton bunch can
be accelerated steadily and the beam quality is well maintained as time goes on.}
\end{figure}

Fig.~\ref{f4} shows the detailed ion acceleration process. As expected, the foil deformation is
avoided effectively. The central part of the foil with the transverse size $\thicksim \sigma_{L}$
can be accelerated as a whole. We again find that the protons move to the front of the heavier carbon
ions at the very beginning. This can be understood by considering the different charge-to-mass ratios of protons
and carbon ions, and the different spatial location of equilibrium for each ion species within their respective effective potentials in the accelerating frame~\cite{s12}. Such a separation is determined by the balance between the inertial and
the electrostatic forces, as shown in Fig.~\ref{f5}(a). However, in comparison with the 1D case we observe additional effects such as the RT-like instability and transverse filamentation which effect the ion acceleration.
As a result of transverse effects, some carbon ions spread into the space trailing the target after $t=22.5T_{0}$ and evolve
into a cloud instead of a compact bunch as seen in Fig.~\ref{f2}(a). However, a dense front always trails the protons.
The sharp front separating the two species is well defined, as shown in Fig.~\ref{f5}(a).

We observe a clear RT-like instability in the 2D simulations. A typical "$\lambda$ structure"~\cite{s10} can be recognized in both the ion and electron density distributions, as shown in Fig.~\ref{f4}. The proton layer is strongly effected, which can be understood by considering
the relativistic laser transparency of the foil,  which occurs when $a > \pi (n_{e}/\gamma n_{c})({l}/{\lambda})$. 
In view of the ultralow thickness of the foil in this case, the electromagnetic wave can partially penetrate the foil and
interact with both ion species as seen in Fig.~\ref{f4}(f). This would seed the RT instability at the proton-carbon interface. Gradually, the protons also suffer from the RT
instability and show an obvious "$\lambda$ structure". However, the trailing carbon ion front contributes to slowing down the local bunching of the
protons and hence suppresses the transverse proton-RT instability. This is the reason why we still
observe a semi stable layer of protons in front of the carbon ions at a later time, i.e., $t=30T_{0}$, though the surface of the layer is
strongly corrugated. Fig.~\ref{f5}(c) shows the ion energy spectra
at t=20, 30, and 40$T_{0}$. Although a mono-energetic peak is observed at the very beginning, it becomes broader after $t=30T_{0}$.
Eventually, the protons also evolve into a cloud in space.

In view of the negative effects of the laser penetration, we propose using a moderate intensity laser pulse or a higher electron density foil to avoid the RT-like instability from effecting the proton layer. We estimate the critical laser intensity for this case by
\begin{equation}
a \sim \pi\frac{n_{e1}}{n_{c}}\frac{l}{\lambda}, n_{e1}=Z_{C}\times n_{C}\label{eq:transparence}
\end{equation}

For comparison, we perform another simulation to show improved proton beam acceleration.

\begin{figure}[!htb]
\suppressfloats\includegraphics[width=8.6cm]{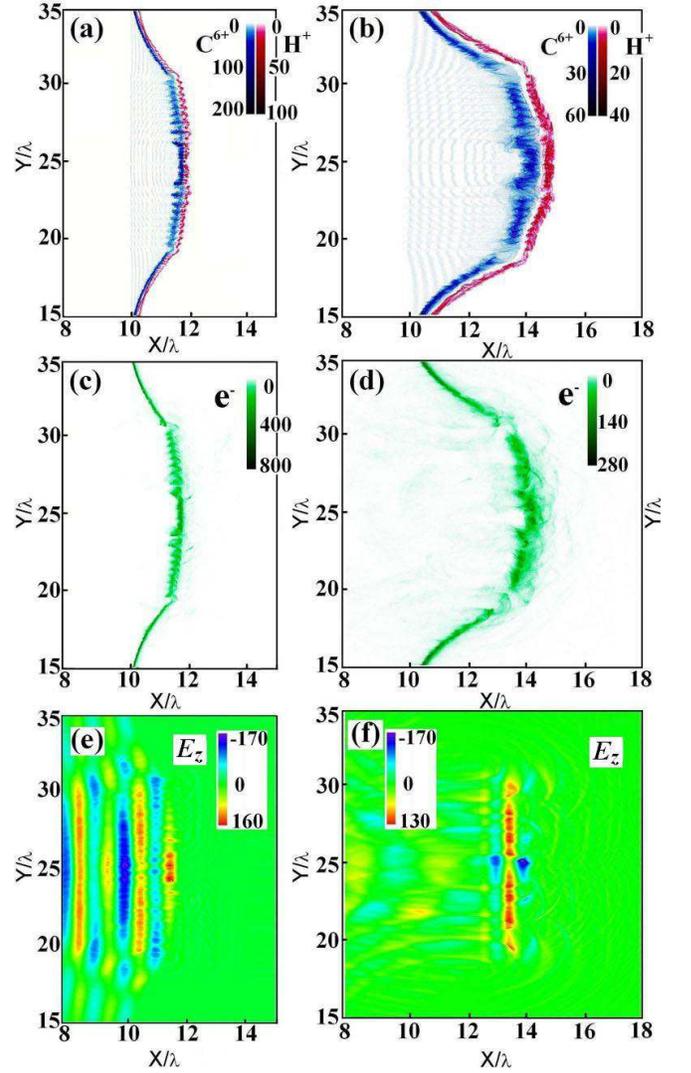}\caption{\label{f6}(Color online)
Density distributions of protons and carbon ions in the case B at (a) $t=17.5T_{0}$ and (b)
$22.5T_{0}$. The proton-RT instability is significantly suppressed as compared with the case A.
The corresponding electron density distribution and the laser electric field $E_{z}$
are shown in frames (c-d) and (e-f), respectively.}
\end{figure}

In case B, the foil is composed of fully ionized carbon ions and protons with number
density ratio $n_{C}:n_{H}=4:1$. The electron density is increased to $n_{e}=500n_{c}$ ($\sim5.5\times10^{23}/cm^{3}$) so that the foil
is opaque to the incident laser pulse. All the other parameters are the same as in the case above. Fig.~\ref{f6}
shows the simulation results. In this case, the foil is well maintained for a much longer time as
compared with the case A. The protons again form a distinct layer from the carbon ions and always ride on the carbon ion front. The laser pulse does not penetrate the foil and the radiation pressure mainly acts on the carbon ion layer,
as depicted in Fig.~\ref{f6}(e-f). As a consequence, the proton layer is kept stable and is less effected by the evolution of the RT instability in the carbon layer.
Compared with case A, the sharp front separating the two species is much more smooth. Most electrons are moving together with the carbon ions, and the spread carbon layer act as a "buffer" for proton acceleration.
This leads to a more stable acceleration structure, which is shown
by the energy spectra in Fig.~\ref{f5}(d). We can see that the energy peak is more pronounced and the spread (FWHM)
is about $20\%$ at $t=30T_{0}$, which is only half of that in the case A. Meanwhile, we find that the cut-off
energy of the carbon ions always peaks the proton spectrum in  the both case A and B. 

It is interesting to note that a similar phenomenon was also observed in Ref.~\cite{s3b,s19}, where mass-limited targets were studied and the results were interpreted as
the direct Coulomb explosion (DCE). In that regime, electrons are overtaking ions as flying compact layers
and the return current is inhibited when mass-limited targets are used. The accelerating field from the Coulomb explosion of the heavier ions accelerates the lighter protons forward.
In our case, however, the laser radiation pressure is dominating the proton acceleration because the electrons always accompany both the carbon ions and protons.
The bulk electrons enable a good screening of the electrostatic field and thus effectively prevent the ions from the Coulomb explosion.
The foil is accelerated forward as a compact dense quasi-neutral plasma until the carbon ions evolve into a cloud in space.
However, we find that the Coulomb explosion of the proton layer becomes significant later. The peak proton energy and the cut-off energy of the carbon ions
continue to increase slowly after the laser-foil interaction ends. After the interaction, i.e., $t=35T_{0}$, the carbon ions evolve into a cloud in space and
the electrons are extensively heated,
and expand with the carbon ions. Finally, the Debye length of the electrons may become larger than the proton layer thickness.
Assuming the thickness of the proton layer being $l_{H,0}$ initially and $l_{H, t}$ at a specific time point, we have $\pi \sigma_{L}^{2}n_{H,0}l_{H,0}=\pi \sigma_{T}^{2}n_{H,t}l_{H,t}$
according to particle number conservation. We can then get a condition for the onset of the Coulomb explosion in this two-ion-species regime as follows

\begin{equation}
T_{e,t}[MeV]>8\times(\frac{n_{H,0}l_{H,0}}{n_{c}\lambda})^{2}\frac{n_{e,t}n_{c}}{n_{H,t}^{2}}\label{eq:CE}
\end{equation}

\noindent where $T_{e,t}$ and $n_{e,t}$ are the electron temperate and density in the proton layer at the time $t$, respectively.
At the post-interaction stage, the electron heating is significant so that the electron temperature is much larger than the value by Eq.~\eqref{eq:CE}, and the Coulomb explosion occurs. For example, at $t=35T_{0}$ in the case B, the proton layer thickness is about $l_{H}=1\mu m$ and the averaged Debye length of the electrons in the proton layer is $d=1.6\mu m$.
In this situation, the effects of the Coulomb explosion should be taken into account~\cite{s222,s223}.
Coulomb explosion is undesirable because it
broadens the final proton energy spectra, though the ion energy can be increased slightly.

It is also necessary to note a recent paper by Qiao, et al.,~\cite{s29}, where they also observed the ion species separation by the laser radiation pressure in the ``leaky light sail'' regime.
However, we did not see such a wide spatial gap between the heavier carbon ions and the light protons. Both the simulations by VLPL and VORPAL~\cite{s30} show almost the same results as elucidated above.

\section{Three-interface model and discussions}

\begin{figure}[!htb]
\suppressfloats\includegraphics[width=7cm]{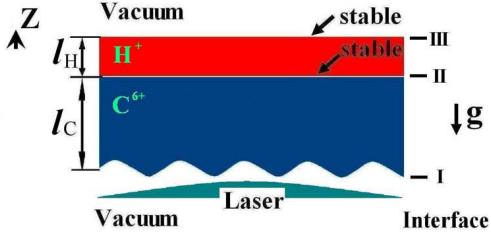}\caption{\label{f7}(Color online)
Schematic of the three-interface model. The top red marks the protons and the bottom blue
shows the carbon ions. The laser pulse is irradiating on the foil from the bottom.
Three interfaces can be recognized in the figure: carbon-vacuum, carbon-proton, and proton-vacuum.}
\end{figure}

The stability of the proton acceleration in the 2D cases above can
be attributed to two crucial effects. First, the protons advance
ahead of the carbon ions and form a thin layer. Such a separation
of the ion species can be understood within the 1D formalism
developed in Sec.(\uppercase\expandafter{\romannumeral2}) and the
1D simulations above. Second, a much thicker and heavier ion layer
trailing behind the proton layer acts as an effective "buffer" to
prevent short-wavelength perturbations from feeding through into
the thin proton shell. As was shown earlier~\cite{s11}, in the
absence of such a buffer (e.g., when carbon ions are either absent
or spatially separated from the protons), the entire proton layer
is affected by the RT-like instability. We introduce a simple
three-interface model, as shown in Fig.~\ref{f7}, to explain the
stabilization of the proton beam acceleration. In our case, both
species have two interfaces: one with vacuum and one with the
other species. The only unstable interface is the carbon-vacuum
boundary, where the laser radiation pressure directly acts on the
carbon plasma. Below we have basic single-fluid hydrodynamic
equations describing the RT instability. Detailed derivation of
these equations  will be presented elsewhere. The following
assumptions are used: (a) electron temperature is much higher that
the ion temperature: $T_{e} \gg T$; (b) plasma ions are fully
neutralized by the plasma electrons; (c) the entire plasma is
accelerated with the constant acceleration $\overline{g}$. Under
these assumptions, the single-fluid momentum equation in the
accelerated in the $z-$ direction reference frame can be written
as:
\begin{equation}
\rho(\frac{\partial \overline{v}}{\partial t} +
\overline{v} \cdot \nabla\overline{v}) =
-\nabla p - \rho \overline{g}, \label{eq:motion}
\end{equation}

\noindent where $\rho$, $\overline{v}$, and $p$
are the mass density, velocity, and ion pressure, respectively.
Linearizing this equation around the plasma equilibrium, and
assuming an isothermal electron response such that $\delta p =
(Z_i/m_i) \left( \delta \rho \right) T_e$, we find that the
perturbed pressure $\delta p$ satisfies
\begin{equation}\label{eq:linearized_motion}
    \rho \frac{\partial \overline{v}}{\partial t} =
    - \overline{\nabla} \delta p - \delta \rho \overline{g},
\end{equation}
which can be even further simplified by assuming incompressibility
of the plasma in response to the RT-like instability:
\begin{equation}
\frac{\partial^2}{\partial z^2}\delta p= k_{RT}^2\delta p\label{eq:pressure}
\end{equation}

\noindent where $k_{RT}$ is the perturbation wave-number in the direction
perpendicular to the acceleration, i.e. parallel to the
interface. Noting that $\delta p$ is discontinuous across the {\it
unperturbed} boundary, we obtain a solution $\delta
p=Ae^{-k_{RT}z}+Be^{k_{RT}z}$ away from interfaces, with A and B
being the amplitude coefficients of the perturbation. These
solutions have to be matched across every interface to satisfy the
following continuity conditions: (i) $v_{z}$ is continuous
(consequence of the single-fluid description); (ii) $\delta p -
i\frac{g}{\omega}\rho_{0}v_{z}$ is continuous (consequence of the
pressure balance across the {\it perturbed} boundary). We can
finally derive from this simple model that the amplitude of the RT
instability is exponentially decaying away from the unstable
interface:

\begin{equation}\label{eq:amplitude}
    \frac{v_z^{H/C}}{v_z^{C/vac}} \sim e^{-k_{RT}l_{i}}
\end{equation}

\noindent where $l_{i}$ is the thickness of the "buffer" ion
layer, and $v_z^{H/C}$ ($v_z^{C/vac}$) are the velocity
perturbations at the proton/carbon (carbon/vacuum) interfaces. In
the case B, we know that $l_{C}$ is a few times larger than
$l_{H}$. In this situation, the long-wavelength perturbation in
the carbon layer should take much more time to grow. Finally, the
perturbation is exponentially attenuated before reaching the
proton layer. We can interpret the carbon ion cloud as a "buffer"
or "cushion" for the proton-RT instability. The carbon ion front
always trails the protons and helps to confine the spatial
development of the proton-RT instability, i.e., local bunching of
the proton beam ("$\lambda$ structure"). Thus the RT-like
instability in the proton layer is significantly suppressed.

\begin{table}[htbp]
\caption{\label{tab:test}Case List}
\begin{tabular}{lccc}
  \toprule
    Label & Composition & $n_{C}$:$n_{H}$ & $n_{e}$ \\
  \midrule
  Case A & $C^{6+}$, $H^{+}$ & 1:1 & 320$n_{c}$\\
  Case B & $C^{6+}$, $H^{+}$ & 4:1 & 500$n_{c}$\\
  Case C & $C^{6+}$, $D^{+}$ & 4:1 & 500$n_{c}$\\
  Case D & $C^{6+}$, $T^{+}$ & 4:1 & 500$n_{c}$\\
  Case E & $H^{+}$ & - & 500$n_{c}$\\
  Case F & $C^{5+}$, $H^{+}$ & 4.8:1 & 500$n_{c}$\\
  Case G & $C^{3+}$, $H^{+}$ & 8:1 & 500$n_{c}$\\
  \bottomrule
\end{tabular}
\end{table}

\begin{figure}[!htb]
\suppressfloats\includegraphics[width=8.6cm]{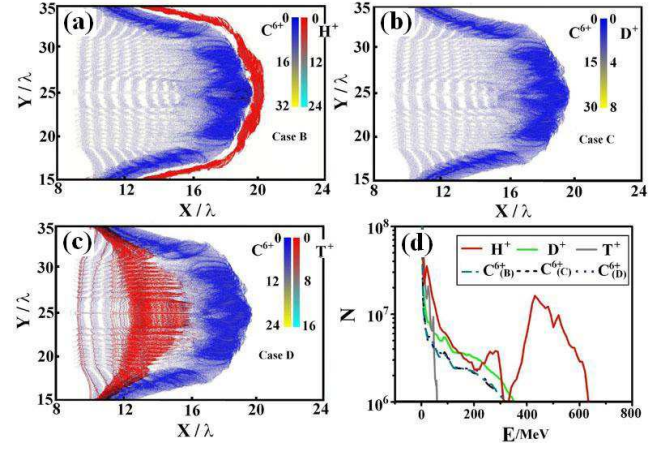}\caption{\label{f8}(Color online)
Ion density distributions in the (a) case B (C-H foil), (b) case C (C-D foil), and (c) case D
(C-T foil) at $t=30T_{0}$. In the C-D foil, both species do not separate from each other.
(d) Ion energy spectra  in the case B, C, and D at $t=40T_{0}$.
Here, C(B), C(C), and C(D) denote the carbon ions in the case B, C, and D, respectively.
A stable acceleration structure exists only in the case B.}
\end{figure}

It is helpful to consider the problem from the view of the
classic RT instability~\cite{s20a,s20b} which occurs when a light
fluid is accelerated into a heavier fluid. In the simulations, the
proton layer is much lighter than the subsequent carbon layer, and
thus proton layer can keep stable. We also perform simulations
where we employ carbon-deuterium (C-D foils, case C,
$Z_{D}/m_{D}=1/2$) and carbon-tritium (C-T foils, case D,
$Z_{T}/m_{T}=1/3$) instead of hydrocarbon (C-H foils, case B,
$Z_{H}/m_{H}=1/1$). All other parameters are the same as in the
case B. The simulation results are depicted in Fig.~\ref{f8}. As
discussed in Sec.(\uppercase\expandafter{\romannumeral2}), the
separation of ion species depends on the ion charge-to-mass ratio.
For the C-D foil, both the carbon ion and the deuterium have the
same $Z_{i}/m_{i}$ and therefore do not separate from each other
(see Fig.~\ref{f8}(b)). The laser pulse pushes both of them
forward together. In this case, it is expected that the RT
instability shall deteriorate both ion energy spectra (see
Fig.~\ref{f8}(d)). For the C-T foil, we do observe ion
separation initially because the carbon ions have a larger
$Z_{i}/m_{i}$ and hence they move to the front of the tritium.
However, the tritium is initially compressed into a very thin
layer, which leads to fast growth of the perturbation at the
vacuum-tritium interface. The instability soon reaches the
tritium-carbon interface and pollutes the front carbon ion beams.
Obviously, such a structure is unstable, and we do not observe a mono-energetic
beam as seen in Fig.~\ref{f8}(d).

\begin{figure}[!htb]
\suppressfloats\includegraphics[width=8.6cm]{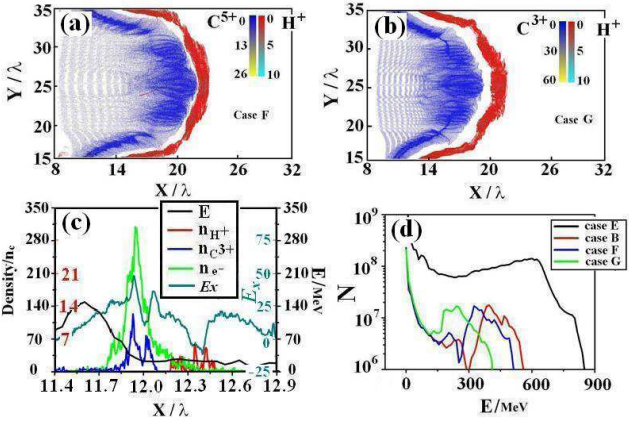}\caption{\label{f9}(Color online) Density distributions of protons and carbon ions ((a) $H^{+}$, case E,(b) $C^{5+}$, case F, and (c) $C^{3+}$, case G.) at $t=35T_{0}$.
(d) Proton energy spectra in the case B, E, F, and G at $t=35T_{0}$.}
\end{figure}

We also investigate effects of the charge state of the carbons on
the stability of the proton acceleration. In the simulations, we
keep the proton density same but vary the charge on carbon ions.
$C^{5+}$ (case F, $Z/m=1/2.4$) and $C^{3+}$ (case G, $Z/m=1/4$)
are taken into account. For comparison, a pure hydrogen SFT (case
E) is also considered. All the other parameters are the same as
that of the case B. The simulation results and the proton energy
spectra are presented in Fig.~\ref{f9}.

It is interesting that with the decrease of the carbon ion charge state, the gap between the carbon ions and protons grows,
which leads to poorer monochromaticity of the proton energy spectrum. It implies that the RT instability is much more severe in the case of the lower charge state of the carbon ions.
This effect is beyond the scope of the current three-interface model and will be studied further. Besides, we believe that the Coulomb explosion of the proton
layer also plays a role in case G because most electrons are with the carbon ions, far away from the protons.
The wide gap prevents the electrons from neutralizing the proton layer. Finally, the electron density becomes very low in the proton layer. According to Eq.~\eqref{eq:CE}, the electron temperature for the onset of the Coulomb explosion is reachable. For example, at $t=20T_{0}$, the proton layer thickness, $l_{H}=0.3\mu m$, is smaller than the averaged Debye
length of the electrons in the proton layer, $\lambda_{D}=0.4\mu m$. As shown in Fig.~\ref{f9}(c), we observe a clear dipped accelerating field with a negative component in the proton layer,
which is a clear signature of Coulomb explosion. Gradually, the protons self-expand in space and the energy spectrum broadens.
This is different from the case B where the Coulomb explosion might occur after the interaction ends.
Additional simulations show that when the charge state of the carbon ions is lower than $3$, both the RT instability and
the Coulomb explosion of the proton layer become violent enough to destroy the mono-energetic proton beam. However, in the relativistic regime, the laser pulse
is powerful enough to fully ionize the carbon atoms. We therefore believe it is possible to observe stable proton acceleration in the in experiments.

Besides, we believe that a smaller transverse size of the foil would benefit the stabilization of the proton acceleration
in this two-ion-species regime. In the simulations, we apply a periodic boundary condition in the transverse direction and use the transverse size of the simulation box $Y=50\lambda$. When we decrease this size to $Y=25\lambda$, we observe considerable suppression of the penetration of the perturbation into the target.

\section{3D simulation results}

\begin{figure}[!htb]
\suppressfloats\includegraphics[width=8.6cm]{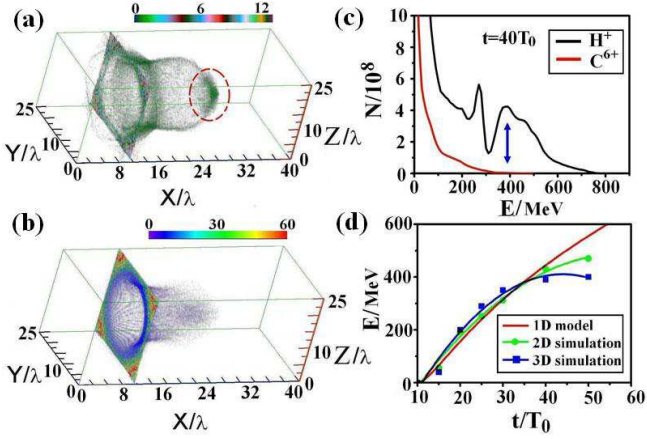}\caption{\label{f10}(Color online)
Density contours of (a) protons and (b) carbon ions in 3D simulations at $t=40T_{0}$. A clear proton bunch is
formed before the carbon ion front. The energy spectra of protons and carbon
ions are shown in the frame (c). The proton peak energy evolution in the 2D case and 3D
case is shown in frame (d). For comparison, the red color marks the theoretical predictions
of the 1D RPA model.}
\end{figure}

Finally, 3D simulations are performed to check the robustness of the stable acceleration mechanism.
To save computational time, the 3D simulation box is $40\lambda\times25\lambda\times25\lambda$,
sampled by $4000\times200\times200$ cells. Each cell contains 27 particles initially. The ultrathin
SFT is composed of fully ionized carbon ions and protons with the density ratio $n_{C}:n_{H}=4:1$. Periodic boundary
conditions for particles and absorbing boundaries for fields are applied. The total laser energy is $\sim500J$ and the duration is $33fs$. All the other parameters are
the same as in the case B above except $\sigma_{T}=5\lambda$ to match the laser focal spot
$\sigma_{L}=6\lambda$. Fig.~\ref{f10}(a,b) illustrates the density distributions of the protons
and the carbon ions at $t=40T_{0}$. We can see a clear compact proton bunch with a few
nano-Coulomb riding on the front of the carbon ions, which agrees well with the 2D simulation results.
The carbon ions spread extensively in space. This should be attributed to
multi-dimensional effects, i.e., the fast growth of the RT instability in the carbon ion layer,
as discussed in Sec.(\uppercase\expandafter{\romannumeral5}).

In Fig.~\ref{f10}(c), we present the ion energy spectra at $t=40T_{0}$. As expected, we observe a clear
mono-energetic peak for the protons. The peak is well maintained for a very long duration even
after the laser-foil interaction ends. On the contrary, the spread carbon ions
show a quasi-exponentially decaying in the energy spectrum. The cut-off energy of the carbon ions
is almost the same as the peak energy of the protons, as observed in the 2D cases. It indicates that the
carbon ion front indeed trails the proton layer during the acceleration. We also
find that there is another small energy peak for protons in the energy spectrum. By analyzing the
ion energy distributions in space, we find that the protons with energy $\sim 280MeV$ are situated in
the vicinity of the carbon ion front. We believe it is due to the incomplete separation of the
protons from the heavier carbon ions in the 3D case. Fig.~\ref{f10}(d) presents the proton peak energy
evolution both in the 2D and 3D simulations. Overall, 3D simulation results fit well with the 1D model
marked by the red curve and the 2D simulations by the green curve. This demonstrates the domination of the RPA mechanism.
The yielded proton beam with the peak energy $\sim 400MeV/u$ may have diverse potential applications
in the future, such as in medical therapy of deep seated tumors and in the development of future
compact ion accelerators.

\section{Conclusions}

In conclusion, we investigate the detailed ion acceleration from an ultra-thin hydrocarbon foil by
use of multi-dimensional PIC simulations. A stable compact proton beam acceleration is observed,
for the first time, in the 3D geometry. This should be attributed to two effects: ion species
separation and heavier ion spreading in space. The laser pulse does not penetrate the foil and
the radiation pressure mainly acts on the carbon ion layer.
The carbon ions act to buffer the compact
proton layer from the RT-like instability. The proposed three-interface model well describes
the simulation results and is further supported by simulations of various compound foils such as C-D, C-T, and pure hydrogen foils.
It is also found that with the decrease of the carbon ions charge state, both the RT instability and the Coulomb explosion become
increasingly violent and tend to degrade the mono-energetic proton beam. Finally,
the robustness of the stable two-ion-species regime is checked by the full 3D simulations.

With the development of the nano-technology~\cite{s21}, polyethylene or mylar foils could be available
soon in experiments.
Compared with the normally used two-layer foils, they are much easier to fabricate.
Benefiting from new state-of-the-art lasers such as HiPER and ELI~\cite{s22}, we believe that the stable
acceleration mechanism described above will be experimentally demonstrated soon and has the potential for applications in science and medicine.

\section*{Acknowledgments}

We thank the fruitful discussions with Dr. N. Kumar and Dr. A. Upadhyay. This work is supported by
the DFG programs GRK1203 and TR18. T.P.Y acknowledges financial support from the China Scholarship
Council and the NSAF program (Grant No. 10976031). G.S. acknowledges the support of the U.S. DOE
Grants No. DE-FG02-05ER54840 and DE-FG02-04ER41321. M.C. acknowledges support from the Alexander von Humboldt Foundation.
The simulations were performed on ATTO in HHUD.


\newpage
\section*{Figure Captions}\suppressfloats

Fig.\ 1. (Color online) (a) Laser intensity evolution. Here, $I=(E_{y}^2+B_{z}^2)/2$ is normalized to
$I_{0}=2.74\times 10^{18} W/cm^{2}$. The laser-foil interaction ends at $t=45T_{0}$.
(b) Carbon ion and proton density distributions at $t=25T_{0}$ and $45T_{0}$.
 The red color shows the accelerating field $E_{x}$ which is normalized to $E_{0}=m_{e}c\omega/e=3.2\times10^{12}V/m$.
(c) Carbon ion and proton phase space distributions at $t=25T_{0}$ and $45T_{0}$.
(d) Proton energy evolution from 1D PIC simulations and 1D RPA model.

Fig.\ 2. (Color online) Energy spectra of (a) carbon ions and (b) protons at $t=35T_{0}$. The ion density ratio
$n_{C}:n_{H}$ in all three cases is 1:1 (case 1), 4:1 (case 2),
and 1:4 (case 3), respectively. A clear mono-energetic peak for both carbon ions
and protons is observed in all three cases.

Fig.\ 3. (Color online) Schematic of the two-ion-spices shaped foil target (SFT). A Gaussian laser pulse is incident
from the left boundary. The profile of the foil is defined by three parameters: $l_{0}$,
$l_{1}$, and $\sigma_{T}$. The foil is composed of mixed carbon ions and protons with
various particle density ratios $n_{C}:n_{H}$.

Fig.\ 4. (Color online) Density distributions of protons and carbon ions in the case A at (a) $t=17.5T_{0}$
and (b) $22.5T_{0}$. RT instability is observed initially for both carbon ions and protons.
The corresponding electron density distribution and the laser electric field $E_{z}$ are
shown in frames (c-d) and (e-f), respectively. Here, the density is normalized to $n_{c}$
and the field to $E_{0}$.

Fig.\ 5. (Color online) (Color online) Particle density distributions and the electron energy distribution as well as the accelerating
  fields $E_{x}$ on the laser axis at $t=20T_{0}$ in the (a) case A and (b) case B.
 Energy spectrum evolutions of the protons and carbon ions at $t=20, 30, 40T_{0}$
 in the (c) case A and (d) case B. For the case B, the compact proton bunch can
be accelerated steadily and the beam quality is well maintained as time goes on.

Fig.\ 6. (Color online) Density distributions of protons and carbon ions in the case B at (a) $t=17.5T_{0}$ and (b)
$22.5T_{0}$. The proton-RT instability is significantly suppressed as compared with the case A.
The corresponding electron density distribution and the laser electric field $E_{z}$
are shown in frames (c-d) and (e-f), respectively.

Fig.\ 7. (Color online) Schematic of the three-interface model. The top red marks the protons and the bottom blue
shows the carbon ions. The laser pulse is irradiating on the foil from the bottom.
Three interfaces can be recognized in the figure: carbon-vacuum, carbon-proton, and proton-vacuum.

Fig.\ 8. (Color online) Ion density distributions in the (a) case B (C-H foil), (b) case C (C-D foil), and (c) case D
(C-T foil) at $t=30T_{0}$. In the C-D foil, both species do not separate from each other.
(d) Ion energy spectra  in the case B, C, and D at $t=40T_{0}$.
Here, C(B), C(C), and C(D) denote the carbon ions in the case B, C, and D, respectively.
A stable acceleration structure exists only in the case B.

Fig.\ 9. (Color online) Density distributions of protons and carbon ions ((a) $H^{+}$, case E,(b) $C^{5+}$, case F, and (c) $C^{3+}$, case G.) at $t=35T_{0}$.
(d) Proton energy spectra in the case B, E, F, and G at $t=35T_{0}$.

Fig.\ 10. (Color online) Density contours of (a) protons and (b) carbon ions in 3D simulations at $t=40T_{0}$. A clear proton bunch is
formed before the carbon ion front. The energy spectra of protons and carbon
ions are shown in the frame (c). The proton peak energy evolution in the 2D case and 3D
case is shown in frame (d). For comparison, the red color marks the theoretical predictions
of the 1D RPA model.

\newpage

\newpage

\newpage

\newpage

\newpage

\newpage

\newpage

\newpage

\newpage

\newpage

\end{document}